\documentclass[%
 reprint,
preprint,  %preprintnumbers,
 amsmath,amssymb,
 aps,
 longbibliography,
 onecolumn
]{revtex4-1}
\makeatletter
\@ifundefined{@parse@version@dash}{%
\def\@parse@version#1{\@parse@version@0#1}
\def\@parse@version@#1/#2/#3#4#5\@nil{%
\@parse@version@dash#1-#2-#3#4\@nil}
\def\@parse@version@dash#1-#2-#3#4#5\@nil{%
\if\relax#2\relax\else#1\fi#2#3#4 }
}{}
\makeatother

\usepackage{graphicx}% Include figure files
\usepackage{dcolumn}% Align table columns on decimal point
\usepackage{bm}% bold math
\usepackage{lineno}
\usepackage[normalem]{ulem}
\usepackage{color}
\usepackage{wrapfig}
\usepackage{hyperref}

\newcommand{\LBNL}{Lawrence Berkeley National Laboratory, Berkeley, CA 94720}
\newcommand{\UCDavis}{University of California, Davis, CA 95616}
\newcommand{\UCLA}{University of California, Los Angeles, CA 90095}
\newcommand{\UCRS}{University of California, Riverside, CA 92521}
\newcommand{\PurdueU}{Purdue University, West Lafayette, IN 47907}
\newcommand{\IndianaU}{Indiana University, Bloomington, IN 47408}
\newcommand{\KSU}{Kent State University, Kent, OH 44242}
\newcommand{\UIC}{University of Illinois at Chicago, Chicago, IL 60607}
\newcommand{\MSU}{Michigan State University, East Lansing, MI 48824}
\newcommand{\Duke}{Duke University, Durham, NC 27708}
\newcommand{\WSU}{Wayne State University, Detroit, MI 48201}
\newcommand{\UIUC}{University of Illinois at Urbana-Champaign, Urbana, IL 61801}
\newcommand{\pepperdine}{Pepperdine University, Malibu, CA 90263, USA}
\newcommand{\NCSU}{North Carolina State University, Raleigh, NC 27695}
\newcommand{\BNL}{Brookhaven National Laboratory, Upton, NY 11973}
\newcommand{\INT}{Institute for Nuclear Theory, University of Washington, Seattle, WA 98195}
\newcommand{\SBU}{Stony Brook University, Stony Brook, NY 11794}
\newcommand{\UNC}{University of North Carolina, Chapel Hill, NC 27599}
\newcommand{\RICE}{Rice University, Houston, TX 77005}
\newcommand{\UH}{University of Houston, Houston, TX 77204}
\newcommand{\OSU}{The Ohio State University, Columbus, OH 43201}
\newcommand{\texasAM}{Texas A\&M University, College Station, Texas 77843}

\newcommand{\snn} {\sqrt{s_{_{\rm NN}}}}

\newcommand{\txt}[1]{\rm{#1}}

\begin{document}

%\linenumbers

\title{QCD Phase Structure and Interactions
at High Baryon Density:\\
Continuation of the BES Physics Program with CBM at FAIR
%--\\
%{\normalsize A White Paper to the U.S.~Department of Energy\\
%National Science Advisory Committee}\\
\vspace{1cm}
}% Force line breaks with \\
%\thanks{A footnote to the article title}%

\def\authspc{\vspace{-0.2cm}}
\def\affspc{\vspace{-0.2cm}}
%\author{US-CBM Collaboration}
\author{\authspc D. Almaalol, M. Hippert, J. Noronha-Hostler, J. Noronha,  and E. Speranza}\affiliation{\affspc\UIUC}
\author{\authspc G. Basar}\affiliation{\affspc\UNC}
\author{\authspc S. Bass}\affiliation{\affspc\Duke}
\author{\authspc D.~Cebra} \affiliation{\affspc\UCDavis}
\author{\authspc V. Dexheimer,  D.~Keane, S. Radhakrishnan, A.I. Sheikh, M. Strickland and C.Y. Tsang }   \affiliation{\affspc\KSU}
\author{\authspc X.~Dong, V.~Koch, G.~Odyniec and N.~Xu} \affiliation{\affspc\LBNL}
\author{\authspc O. Evdokimov, D. Hofman, M. Stephanov, G.~Wilks and Z.Y.~Ye} \affiliation{\affspc\UIC}
\author{\authspc F. Geurts} \affiliation{\affspc\RICE}
\author{\authspc H.Z.~Huang and G.~Wang} \affiliation{\affspc\UCLA}
\author{\authspc J.Y.~Jia}   \affiliation{\affspc\SBU}
\author{\authspc H.S.~Li and F.Q.~Wang} \affiliation{\affspc\PurdueU}
\author{\authspc J.F.~Liao}   \affiliation{\affspc\IndianaU}
\author{\authspc M.~Lisa}   \affiliation{\affspc\OSU}
\author{\authspc L. McLerran and A. Sorensen}    \affiliation{\affspc\INT}
\author{\authspc C.~Plumberg}   \affiliation{\affspc\pepperdine}
\author{\authspc S. Mukherjee, R. Pisarski, B. Schenke and Z.B. Xu} \affiliation{\affspc\BNL}
\author{\authspc S. Pratt}   \affiliation{\affspc\MSU}
\author{\authspc R. Rapp}    \affiliation{\affspc\texasAM}
\author{\authspc C. Ratti and V. Vovchenko} \affiliation{\affspc\UH}
\author{\authspc T. Sch\"afer} \affiliation{\affspc\NCSU}
\author{\authspc R. Seto}   \affiliation{\affspc\UCRS}
\author{\authspc C. Shen}   \affiliation{\affspc\WSU}

%\author{(for the US--CBM Collaboration)} 
%\affiliation{\LBNL}

\date{\today}% It is always \today, today,
             %  but any date may be explicitly specified
             
\renewcommand\abstractname{{\bf Executive Summary}}
\begin{abstract}
\textbf{We advocate for an active US participation in the international collaboration of the CBM experiment that will allow the US nuclear physics program to build on its successful exploration of the QCD phase diagram, use the expertise gained at RHIC to make complementary measurements at FAIR, and contribute to achieving the scientific goals of the beam energy scan (BES) program.}

Below, we list ways in which the CBM experiment, designed to operate at a high ion luminosity in the beam energy range corresponding to $\snn=2.9$--$4.9\ \txt{GeV}$, will add to measurements started wihin the RHIC BES program:

\begin{itemize}

    \item 
    The RHIC BES program measurements in the energy range $\snn = 7.7$--$54.4\ \txt{GeV}$ show an intriguing non-monotonic behavior of the fourth-order net-proton cumulant as a function of the collision energy, consistent with expectations for a system evolving in the vicinity of a QCD critical point. More statistics and measurements at lower energies are necessary to firmly establish the existence of the critical point and study it in detail. 
    However, with the exception of collisions at $\snn = 3.0\ \txt{GeV}$, the BES fixed-target measurements do not provide the rapidity coverage nor the event statistics needed for cumulant analyses. The CBM experiment is a unique high-intensity experiment with the capability of taking data at event rates up to 10~MHz, offering both the necessary statistics and the rapidity coverage. We propose to perform net-proton cumulant measurements that would bridge the gap between high-statistics measurements from BES and help to decisively answer questions about the existence of the QCD critical point in the regions of the phase diagram accessible by terrestrial experiments.

    \item Studies probing the hadron-QGP phase transition and constraining hypernuclear interactions rely on high-statistics observables. The CBM experiment will provide unprecedented large data sets allowing one to measure a number of such signature observables. We propose to carry out measurements of dilepton spectra, multi-strange hyperons and hypernuclei, and polarization and spin alignment. %, with their dependence on the collided species. 
    These measurements will explore characteristics of dense baryonic matter produced in the collisions, including properties relevant to our understanding of the inner structure of compact stars and dynamics of neutron star mergers.
    %the possible existence of quarkyonic matter and pion/kaon condensates, quantum pion liquids,

    \item %RHIC has been designed to facilitate collisions of gold nuclei. 
    FAIR is able to collide a variety of nuclei, including isobaric species with varying isospin content, which will enable a precise determination of properties of dense QCD matter not yet explored in the RHIC BES. In particular, studies of dense matter up to several times nuclear matter saturation density with varying isospin content could provide crucial constraints for both the nearly-symmetric and asymmetric equation of state, with the latter having particular significance for neutron stars.
    %controlled terrestrial experiments
    We propose high-quality and high-statistics measurements of spectra, flow, and femtoscopic correlations from a systematic scan of beam energies and target--projectile combinations at FAIR/CBM that will shed more light on the dynamics and the equation of state of dense nuclear matter, including potentially measuring the speed of sound in nuclear matter over a broad temperature and chemical potential domain and helping constrain the maximum mass of neutron stars.

\end{itemize}

These high-quality measurements in the FAIR/CBM energy region will build on the emerging success of studies pioneered at RHIC, help answer %many 
some of the key physics questions that originated the BES campaign, and allow reaching the full potential of the scientific program started at RHIC. There is no doubt that the proposed CBM physics program will be essential to addressing questions concerning the phase structure of nuclear matter at the highest baryon densities achievable in a controlled experiment. The US participation in CBM will not only greatly enhance the CBM physics program, but will also strengthen US leadership in nuclear physics. 

\end{abstract}
%\keywords{Suggested keywords}%Use showkeys class option if keyword
                              %display desired
\maketitle
\tableofcontents

%\vfill{}

\section{Physics at high baryon densities}

The RHIC physics program has had defining contributions in the exploration of QCD matter, including the discovery of the Quark-Gluon Plasma (QGP) %A new form of matter, the Quark-Gluon Plasma (QGP), has been discovered 
in very high-energy nuclear collisions %in the early 2000's %at RHIC
~\cite{Adams:2005dq,Adcox:2004mh,Arsene:2004fa,Back:2004je}. For these high-energy collisions at high temperature and vanishing net-baryon density, the QGP-hadron transition is understood to be a smooth crossover based on lattice QCD calculations~\cite{Aoki:2006we}. Since making these discoveries in the early 2000's, scientists have been asking: {\bf What is the structure of the QCD phase diagram in the high net-baryon density ($n_B$) region, or equivalently in the large baryon chemical potential ($\mu_B$) region?}

Investigation of the QCD phase diagram at finite values of the baryon chemical potential and temperature motivated the Beam Energy Scan (BES) physics program, initiated in 2008 and colliding heavy nuclei at energies lower than the top energy ($\snn = 200\ \txt{GeV}$) with the goals of searching for the onset of the QCD phase transition, putting constraints on the nuclear matter equation of state at high baryon density, and extraction of the hyperon-nucleon interaction. The data-taking for the BES program has been successfully completed in 2021, and the corresponding analyses have already started to shed light on the properties of QCD matter at high densities.

\begin{figure}[h]
  \begin{center}
    \includegraphics[width=0.75\textwidth]{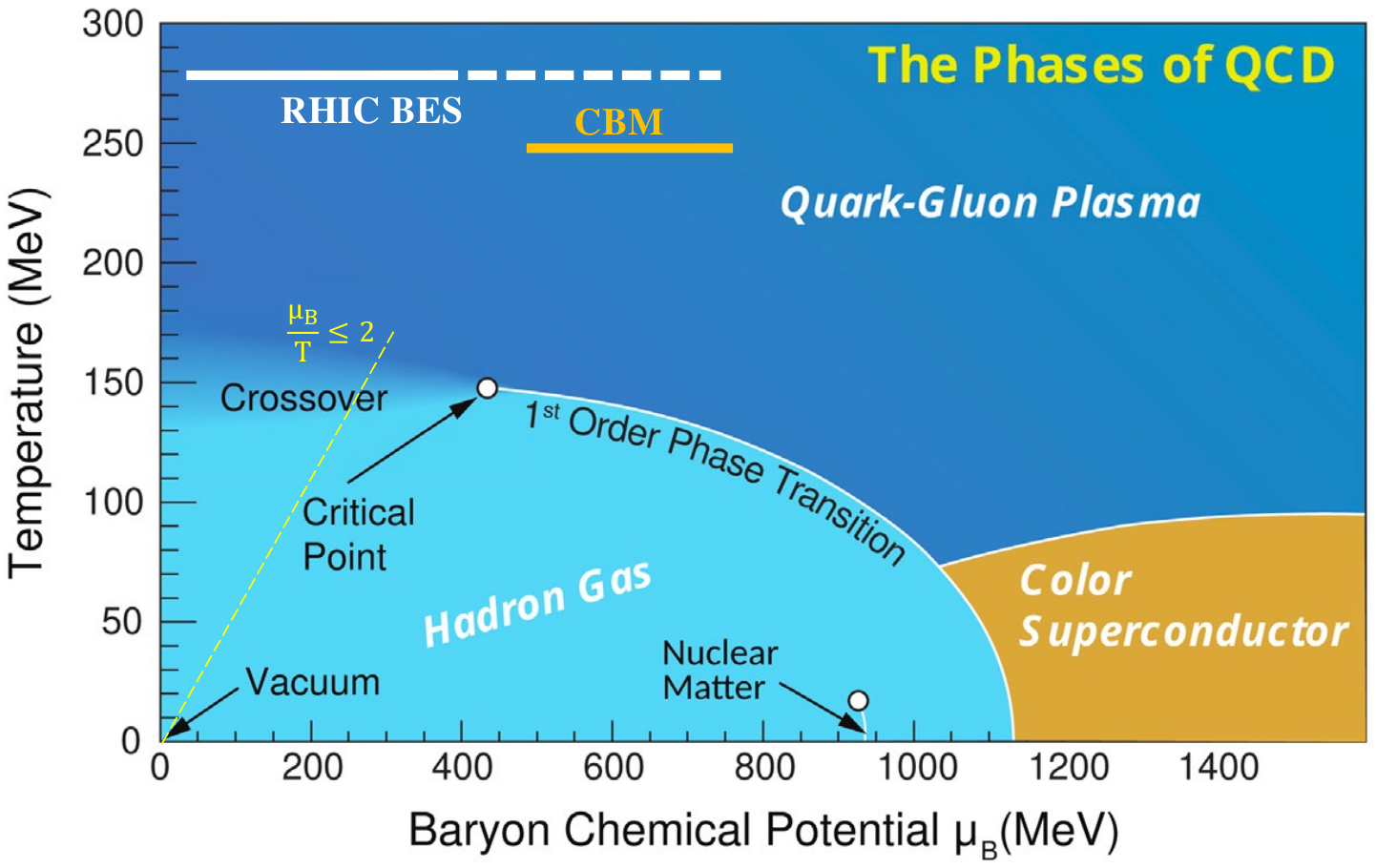}
  \end{center}
   \vspace{-2.cm}
\linespread{1.0}\selectfont{} 
    \caption{Sketch of the QCD phase diagram, incorporating a conjectured QCD critical %end
    point  and the corresponding first-order phase transition line. The solid yellow line indicates the region up to $\mu_B/T \leq 2$ where lattice QCD calculations, predicting a smooth crossover, are valid. 
    The coverage of the RHIC BES program in the collider and fixed target mode (solid and dashed white bars, respectively) and of the future CBM Experiment at FAIR (solid orange bar) are indicated near the top of the figure.
    }
\label{fig1:phasestructure} 
\end{figure}

Model studies indicate that a first-order phase boundary is expected at large $\mu_B$, as illustrated by the white line in Fig.~\ref{fig1:phasestructure}, showing the QCD phase diagram; see discussions in Refs.~\cite{Fukushima:2013rx,Baym:2017whm,Bzdak:2019pkr}. %,US:NPLRP2015}
If this is the case, then there is a QCD critical point (strictly speaking, a critical region for finite systems) between the first-order phase boundary and the smooth crossover indicated by lattice QCD~\cite{Aoki:2006we}. One of the most prominent goals of the BES program is to search for this critical point and the corresponding softening in the equation of state due to a first-order phase transition~\cite{STAR:BESII2014,US:NPLRP2015}.

%-- Fig.1 --===========================================================
%\begin{wrapfigure}{r}{0.5\textwidth}

Nuclear collisions at BES energies, where in every collision a considerable number of hyperons are produced %per collision 
on a scale of several femtometers, present a unique opportunity to study hyperon-hyperon interactions, which are otherwise inaccessible due to their short lifetimes. Hyperon-nucleon and hyperon-hyperon interactions are fundamental ingredients to understand QCD and the equation of state (EoS) that governs the properties of nuclear matter and astrophysical objects such as neutron stars~\cite{Lonardoni:2014bwa}.

General arguments suggest that at nonzero quark density there is a quarkyonic phase, that is a phase of strongly-interacting matter which can be
chirally symmetric yet confined~\cite{McLerran:2007qj}. Such a phase can contain both pion/kaon and quarkyonic condensates~\cite{Pisarski:2018bct}, as well as exhibit behavior of a quantum pion liquid~\cite{Pisarski:2020dnx}. Distinctive signatures of these exotic phenomena are predicted to be accessible in heavy-ion collisions through two-pion correlations~\cite{Pisarski:2021qof} and dilepton spectra~\cite{Hohler:2013eba,Glozman:2022lda}.

We anticipate the BES measurements to yield cutting-edge contributions toward resolving the physics questions described above, and more. At the same time, we are also aware of unique opportunities arising at facilities worldwide, where experiments with different detector and accelerator capabilities are currently performed or constructed. These include the Compressed Baryonic Matter (CBM) experiment at the Facility for Antiproton and Ion Research (FAIR), Germany, which is designed to operate at high ion luminosity in the beam energy range corresponding to $\snn = 2.9$--$4.9\ \txt{GeV}$ and which will probe strongly-interacting QCD matter at baryon chemical potentials $\mu_B \simeq 500$--$800\ \txt{MeV}$. The CBM experiment, with an unprecedented capability on interaction rate, full mid-rapidity coverage, and a possibility to use different colliding species, will allow measurements of observables inaccessible at RHIC.

\section{Results from the RHIC Beam Energy Scan Program}

The Beam Energy Scan program at RHIC covered nucleon-nucleon center-of-mass energies from $\snn = 3$~GeV to 200~GeV, with the corresponding baryon chemical potential ranging from $\mu_B=760$~MeV to 25~MeV; see Table~\ref{tab1}. In particular, during the second phase of the program (BES-II), the STAR experiment took data in the fixed-target mode from $\snn = 3$ to $13.7\ \txt{GeV}$. While BES achieved unprecedented statistics, %However, 
among the fixed target energies only the $\snn = 3\ \txt{GeV}$ data have full mid-rapidity coverage for protons ($|y_p - y_{\txt{beam}}|\leq 0.5$) and very high statistics, which is crucial for most of the physics observables discussed below. 

%%%%%%%%%%%%%%% STAR BES-I and -II energy et al. table %%%%%%%%%%%%%%%%%%
\begin{table}[thb]
    \centering
    \linespread{1.0}\selectfont{} 
        \caption{The RHIC BES program at the STAR detector: nucleon-nucleon center-of-mass energies ($\snn$), collected numbers of events (the second number, if present, refers to BES-II), the year data were produced, the corresponding  center-of-mass rapidity, and the freeze-out parameters in central Au+Au collisions. The collider mode covered the center-of-mass energies $\snn = 7.7$--$200\ \txt{GeV}$, while the fixed-target (FXT) mode covered $\snn = 3.0$--$13.7\ \txt{GeV}$. Among the FXT data, only $\snn = 3\ \txt{GeV}$ data have full mid-rapidity coverage for protons ($|y_p|< 0.5$ and $0.4<p_T<2$ GeV/c).}
    \label{tab1}
    \def\spc{0.16cm}
    \begin{tabular}{cccccc}\hline
        \hspace{\spc}$\snn$ (GeV)\hspace{\spc} & \hspace{\spc}Events ($10^6$)\hspace{\spc} & \hspace{\spc}Year of Data\hspace{\spc} & \hspace{\spc} Rapidity\hspace{\spc} & \hspace{\spc}$\mu_B$ (MeV)\hspace{\spc} & \hspace{\spc}$T$ (MeV)\hspace*{\spc}\\ \hline
        200  & 238     & 2010      & $|y|<$ 0.5    & 25  & 166 \\ %\hline
        64.4 & 46      & 2010      & $|y|<$ 0.5    &  73  & 165 \\ %\hline
        54.4 & 1200    & 2017      & $|y|<$ 0.5    &  83  & 165 \\ %\hline
        39   & 86      & 2010      & $|y|<$ 0.5    &  112 & 164 \\ %\hline
        27   & 30/560  & 2011/2018 & $|y|<$ 0.5    &  156 & 162 \\ %\hline
        19.6 & 15/538  & 2011/2019 & $|y|<$ 0.75    &  206 & 160 \\ %\hline
        17.3 & 560     & 2021      & $|y|<$ 0.75    & 227 & 158 \\ %\hline
        14.6 & 13/325  & 2014/2019 & $|y|<$ 0.75    & 264 & 156 \\ %\hline
        11.5 & 7/230   & 2010/2020 & $|y|<$ 0.75    & 315 & 152 \\ %\hline
        9.2  & 0.3/160 & 2008/2020 & $|y|<$ 0.75    & 355 & 140 \\ %\hline
        7.7  & 4/100   & 2010/2021 & $|y|<$ 0.75    & 420 & 140 \\ 
    %    \hline%\hline
    %    13.7 (FXT)& ?   &    2021   & 2.69 & 276 & ?  \\
    %    11.5 (FXT)& ?   &    2021   & 2.51 & 316 & ?  \\
    %    9.1 (FXT)& ?    &    2021   & 2.28 & 372 & ?  \\
    %    7.7 (FXT)& ?    &    2020   & 2.10 & 420 & ?  \\
    %    7.2 (FXT)& ?    &    2018   & 2.02 & 443 & ?  \\
    %    6.2 (FXT)& ?    &    2020   & 1.87 & 487 & ?  \\
    %    5.2 (FXT)& ?    &    2020   & 1.68 & 541 & ?  \\
    %    4.5 (FXT)& ?    &    2020   & 1.52 & 589 & ?  \\
    %    3.9 (FXT)& ?    &    2020   & 1.37 & 633 & ?  \\
    %    3.5 (FXT)& 120  &    2020   & 1.25 & 670 & 80  \\
    %    3.2 (FXT)& 200  &    2019   & 1.13 & 700 & 80  \\ 
        3.0 (FXT)& 2000 &    2018/2021   & $-1.05< y < 0.5$ & 760 & 80  \\ \hline
    \end{tabular}
\end{table}
%%%%%%%%%%%%%%%%%%%%%%%%%%%%%%%%%%%%%

\subsection{Collective Flow to Probe the Equation of State}\label{sec:collect}

\begin{figure}[t]
\begin{minipage}{0.7\textwidth}
    \hspace{-10mm}
    \includegraphics[width=\columnwidth]{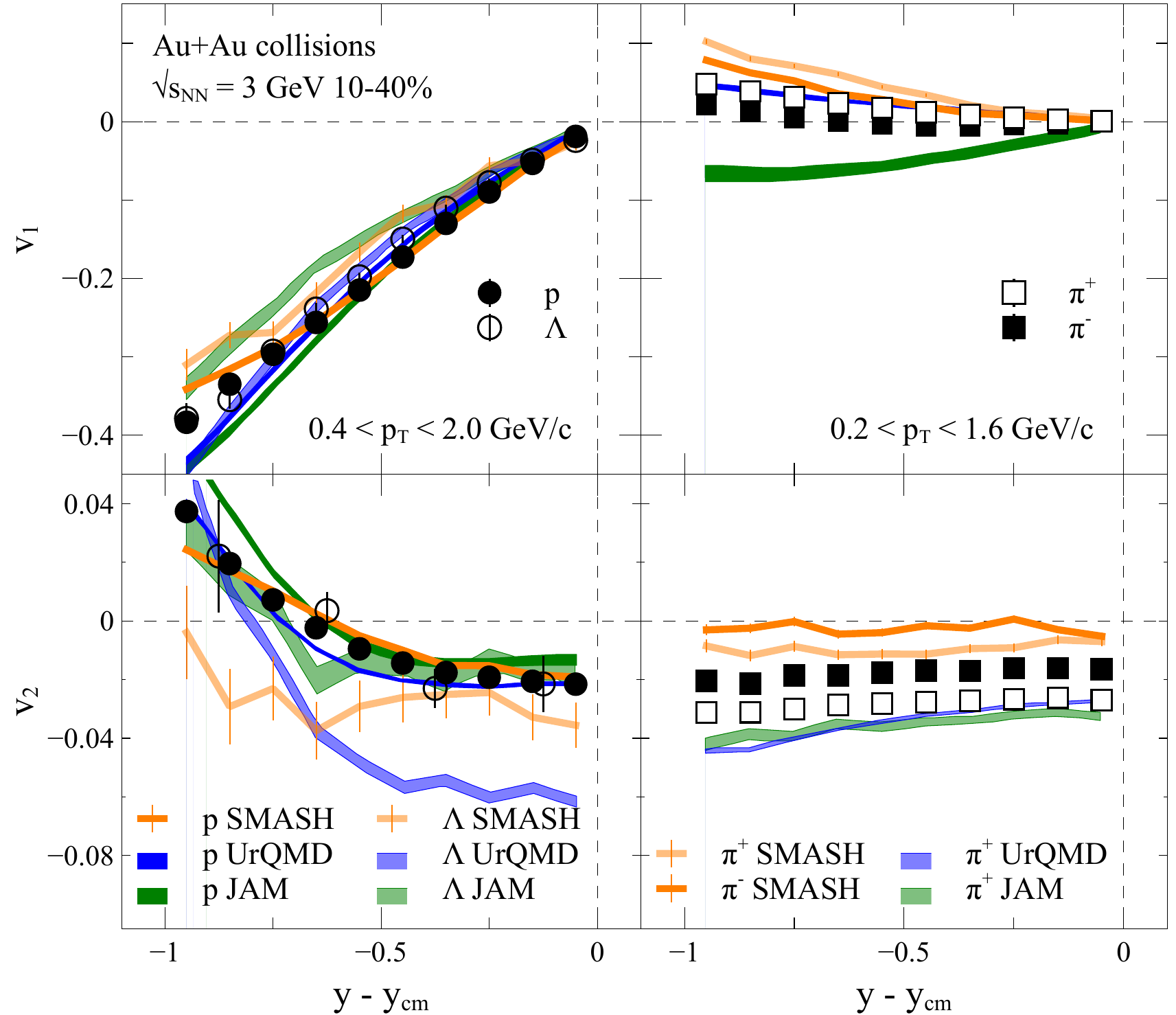}
\end{minipage}
\begin{minipage}{0.28\textwidth}
\vspace{-5mm}
\linespread{1.0}\selectfont{} 
\caption{Rapidity ($y$) dependence of the directed ($v_1$) and elliptic ($v_2$) flow of protons and  Lambda baryons (left panels) and pions (right panels) in $\snn=3.0\ \txt{GeV}$ Au+Au collisions at 10--40\% centrality from STAR measurements \cite{STAR:2021yiu} and UrQMD, JAM, and SMASH hadronic transport simulations \cite{STAR:2021yiu,Oliinychenko:2022uvy}. All simulations used a relatively hard EoS at moderate densities, with the EoS used in SMASH becoming significantly softer at high densities; see \cite{STAR:2021yiu,Oliinychenko:2022uvy} and references therein for more details.}
\label{fig2:flow_3GeV}
\end{minipage}
\end{figure}

Angular distributions of the final state hadrons $dN/d\phi$ %, where $\phi$ denotes the azimuthal angle measured with respect to the reaction plane, 
are highly sensitive to the EoS of symmetric nuclear matter, as shown in multiple hydrodynamic \cite{Stoecker:1980vf,Ollitrault:1992bk,Rischke:1995pe,Stoecker:2004qu,Brachmann:1999xt,Csernai:1999nf,Ivanov:2014ioa,Pratt:2015zsa,Alba:2017hhe,Spieles:2020zaa,Monnai:2021kgu} and hadronic transport \cite{Hartnack:1994ce,Li:1998ze,Danielewicz:2002pu,LeFevre:2015paj,Wang:2018hsw,Nara:2021fuu,Oliinychenko:2022uvy,Steinheimer:2022gqb} studies. For collisions at high baryon densities (i.e., in the BES FXT and CBM energy range), the interplay between the expansion of the fireball and the influence of the spectators is highly dependent on baryonic interactions. Therefore, measurements of flow observables (in particular, the elliptic flow $v_2$ and the slope of the directed flow $dv_1/dy |_{y=0}$) can lead to significant constraints on the high-density EoS, with collisions at $\snn \sim 3$--4 GeV and $\snn \sim 4$--5 GeV being most sensitive to the EoS at densities 2--3 and 3--4 times the saturation density, respectively \cite{Oliinychenko:2022uvy}. Recent studies \cite{Oliinychenko:2022uvy,Steinheimer:2022gqb} including the latest STAR measurements at $\snn = 3.0$ and $4.5$ GeV \cite{STAR:2020dav,STAR:2021yiu} indicate that heavy-ion data are consistent with a relatively hard EoS at moderate densities (1--3 times saturation density) and a relatively soft EoS at high densities (3--5 times saturation density). This creates some tension with neutron star data, which for these density ranges indicate a steep rise in the speed of sound in asymmetric nuclear matter \cite{Bedaque:2014sqa,Tews:2018kmu,Tan:2020ics}. Moreover, while proton flow is relatively well described by models, pion flow (sensitive to the isospin dependence of the EoS, see, e.g., \cite{Liu:2019ags}) and Lambda baryon flow (sensitive to strange interactions, see, e.g., \cite{Li:1996ju,Wang:1998ew,Ko:2000cd}) are not well described, see Fig.\ \ref{fig2:flow_3GeV}.

Precise measurements of differential flow observables will be necessary for more accurate model comparisons. The CBM experiment, with its excellent acceptance, will enable detailed studies of the rapidity-dependence of flow, leading to more meaningful constraints on the symmetric nuclear matter EoS. The unprecedented interaction rate at CBM will also allow to put tighter constraints on hyperon interactions (see Section \ref{sec:hyper} for more details). Moreover, while the RHIC BES program was largely based on a single collision system, Au+Au, at FAIR collisions of a variety of nuclei of different size and isospin content can be achieved. Crucially, a comparison of isobar species with the same mass numbers but different nuclear shape and/or isospin content~\cite{FOPI:1999gxl,FOPI:2007gvb,STAR:2021mii} provides a lever arm to vary the initial condition and observe its importance for the final state~\cite{Xu:2021uar,Zhang:2021kxj,Nijs:2021kvn,Zhang:2022fou,Jia:2022qgl}, as well as allows one to the extract information about the isospin dependence of the EoS~\cite{Xu:2021vpn,Jia:2021oyt}. A scan of nuclei with different mass numbers could also provide insights into the onset of the hadron-quark transition while minimizing the impact of volume fluctuations~\cite{Luo:2013bmi,Braun-Munzinger:2020jbk} and baryon conservation~\cite{Bzdak:2019pkr}, complementing corresponding upcoming studies within the RHIC BES program.

%-- Fig.2 --===========================================================
%\begin{wrapfigure}{r}{0.5\textwidth}
%\vspace{-0.5cm}

While flow measurements capture the macroscopic evolution of systems created in heavy-ion collisions, they may help establish whether the microscopic degrees of freedom at small collision energies ($\snn\sim2.4$--7.7~GeV) are those of deconfined quarks and gluons or confined hadrons. 
%From the microscopic physics perspective, it is not known at this point whether the main degrees of freedom throughout the evolution of systems created at small collision energies ($\snn\sim2.4$--7.7~GeV) are those of deconfined quarks and gluons or confined hadrons. 
For beam energies in the range $\snn \sim 7.7$--$200\ \txt{GeV}$, a hybrid theoretical approach utilizing relativistic viscous hydrodynamics of an almost-perfect fluid (suggesting the relevance of strongly-interacting quark and gluon degrees of freedom) coupled with a hadronic afterburner works remarkably well (with some recent studies suggesting a reasonable description even down to $\snn \sim2.4$--4.3 GeV \cite{Schafer:2021csj,Inghirami:2022afu,Spieles:2020zaa}). On the other hand, at low energies ($\snn \sim 2.0$--$7.7\ \txt{GeV}$) hadronic transport, evolving hadronic degrees of freedom, provides a good description of experimental data \cite{TMEP:2022xjg,Mohs:2020awg,Nara:2021fuu}. %At the same time, recent measurements by the HADES experiment \cite{HADES:2019auv} at $\snn = 2.4$ GeV  indicate that the average temperature of the medium created in these collisions is significantly larger than initially anticipated, leading one to wonder whether the system approaches a deconfined state of matter. 
This raises the question of how the dominant degrees of freedom in heavy-ion collisions change with the beam energy. High-statistics measurements from the CBM experiment, including collective flow measurements and their dependence on the transverse momentum $p_T$, rapidity $y$, and particle species, will provide stringent tests on the modeling approaches and may help determine the correct dynamical description in terms of the microscopic degrees of freedom.

Altogether, the FAIR system scan program will provide an unprecedented diverse set of flow data necessary for understanding the EoS of symmetric and asymmetric dense nuclear matter, the initial state of heavy-ion collisions, and QCD dynamics, complementing the flow measurements performed at RHIC.

\subsection{Net-Proton Fluctuations to Search for the QCD Critical Point}\label{sec:kurtosis}

%-- Fig.3 --===========================================================
%\begin{wrapfigure}{r}{0.5\textwidth}
%\vspace{-0.5cm}
\begin{figure}[htb]
\begin{minipage}{0.59\textwidth}
    \includegraphics[width=\columnwidth]{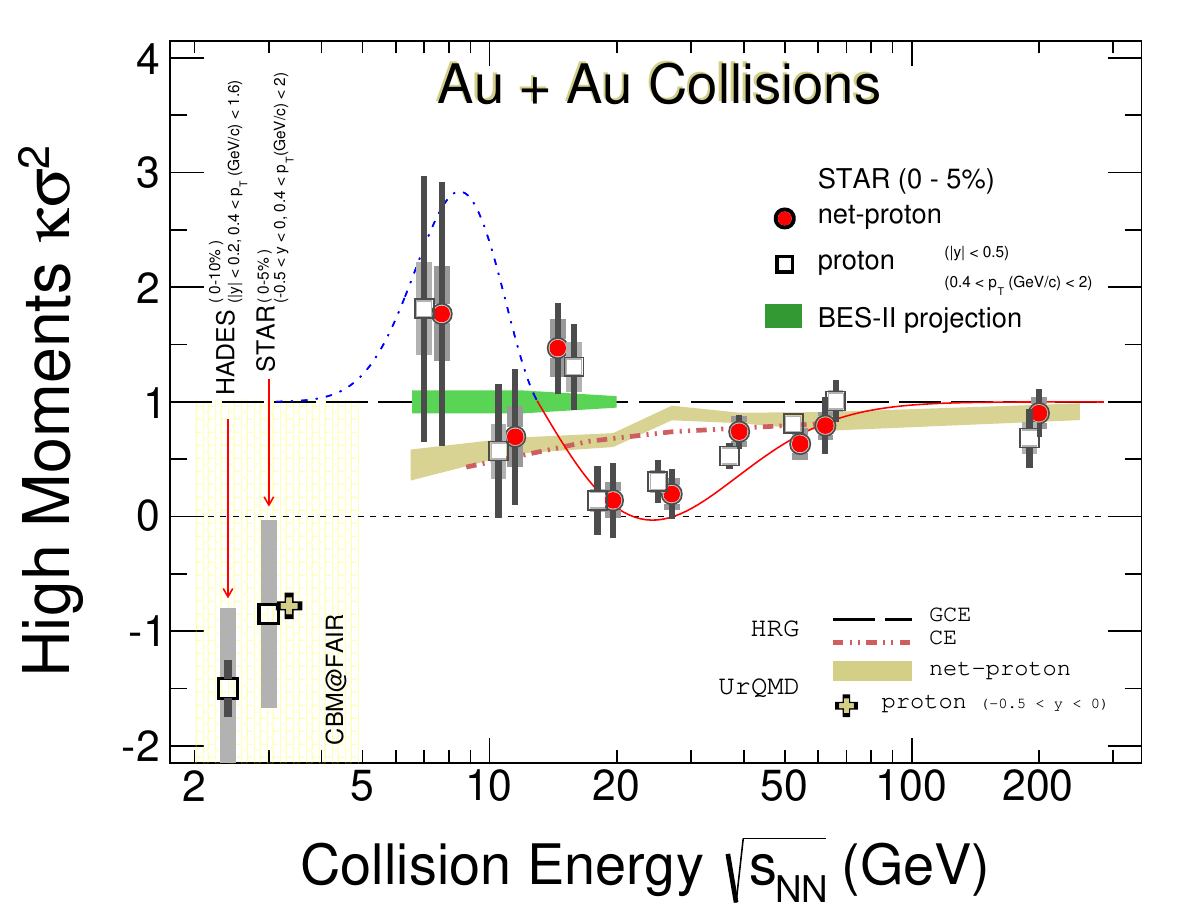}
\end{minipage}
\begin{minipage}{0.4\textwidth}
\vspace{-5mm}
\linespread{1.0}\selectfont{} 
\caption{Energy dependence of the net-proton (filled circles) and proton (open squares) high moments from central Au+Au collisions at RHIC BES~\cite{STAR:2021iop,Adamczewski-Musch:2020slf,STAR:2021fge}. For a comparison with a non-critical baseline, model results from HRG, based on both Canonical Ensemble (CE) and Grand-Canonical Ensemble (GCE)~\cite{Braun-Munzinger:2020jbk}, and transport model UrQMD (cascade mode)~\cite{Bleicher:1999xi,Bass:1998ca} are also shown. The energy range covered by the CBM experiment is shown as the yellow hatched area. The non-monotonic curve indicates the qualitative shape expected due to the contribution of critical fluctuations.}
\label{fig2:net-p}
\end{minipage}
\end{figure}

Figure~\ref{fig2:net-p} shows recent results on the fourth-order net-proton (filled red circles) and proton (open squares) high moments in central Au+Au collisions from %the first phase of \as{not only the first phase as you show the 3.0 GeV point}
the RHIC BES program and the HADES experiment~\cite{STAR:2020tga,STAR:2021fge,HADES:2020wpc} together with model comparisons. The thin solid red and dot-dashed blue lines depict a qualitative prediction for the behavior of the fourth-order net-proton cumulant due to an evolution in the vicinity of a critical region~\cite{Stephanov:2011pb}; the locations of the peak of the dot-dashed blue curve and the dip of the solid red curve %in the figure 
are chosen only for a qualitative comparison, and in particular the could occur at lower collision energies probed at FAIR. 
Intriguing non-monotonic behavior is observed in the data, and while large error bars prevent one from making a more decisive conclusion, the behavior of the fourth-order net-proton cumulant in the energy range $\snn = 7.7$--$27\ \txt{GeV}$ seems to significantly deviate from the non-critical baseline provided by models. In particular, both the hadronic transport model UrQMD (cascade mode)~\cite{Bass:1998ca,Bleicher:1999xi} (gold band) and a thermal model in the canonical ensemble~\cite{Braun-Munzinger:2020jbk} (dot-dashed red line) predict, for a decreasing collision energy, a monotonic suppression in the fourth-order moments due to baryon number conservation, in contrast to the behavior tentatively seen in data. In addition to the fourth-order moments, the experimental data at $\snn \lesssim 20\ \txt{GeV}$ indicate an excess of two-proton correlations as compared to a non-critical baseline including effects due to, e.g., baryon number conservation~\cite{Vovchenko:2021kxx}, with the largest deviations seen at $\snn = 7.7\ \txt{GeV}$. In low-energy collisions, net-proton cumulants have been also recently proposed as a means for extracting the speed of sound and its logarithmic derivative \cite{Sorensen:2021zme}.

Significantly improved statistical precision (and likely reduced systematic uncertainties) are expected from RHIC BES II, as indicated by the green band in Figure~\ref{fig2:net-p}; indeed, this can be already seen from the $\snn = 3.0\ \txt{GeV}$ data point, obtained within an initial analysis utilizing only a fraction of all currently available data. Moreover, the experimental results at $\snn =3\ \txt{GeV}$ are consistent with both the thermal model and UrQMD calculations \cite{STAR:2021fge}, implying that the system at this energy is dominated by hadronic %(dominant by baryonic) 
interactions. 

Different possible explanations of the behavior of the fourth-order moments as a function of the beam energy, including that of the QCD critical point in the baryon-rich regime, have been discussed, however, further measurements at collision energies $\snn = 3.0$--$7.7\ \txt{GeV}$ are required for clear conclusions. This may prove to be challenging at higher fixed-target energies at RHIC, where the kinematics of the collisions together with the geometry of the STAR experiment detector will not allow for measuring higher order moments with a full mid-rapidity coverage, necessary to make meaningful comparisons with the collider mode data. On the other hand, the CBM experiment ~\cite{Friman:2011zz} at FAIR (the energy coverage of which is indicated in Figure~\ref{fig2:net-p} by the yellow hatched area) will enable measurements of higher order moments with the full mid-rapidity coverage required to further explore net-proton fluctuations at high baryon density and decisively establish whether the QCD critical point is located in the region of the QCD phase diagram accessible to terrestrial experiments.

%--==============================================================================
\subsection{Hyperons and Hypernuclei to Probe Baryonic Interactions}\label{sec:hyper}

\begin{figure}[htb]
\includegraphics[width=0.95\columnwidth]{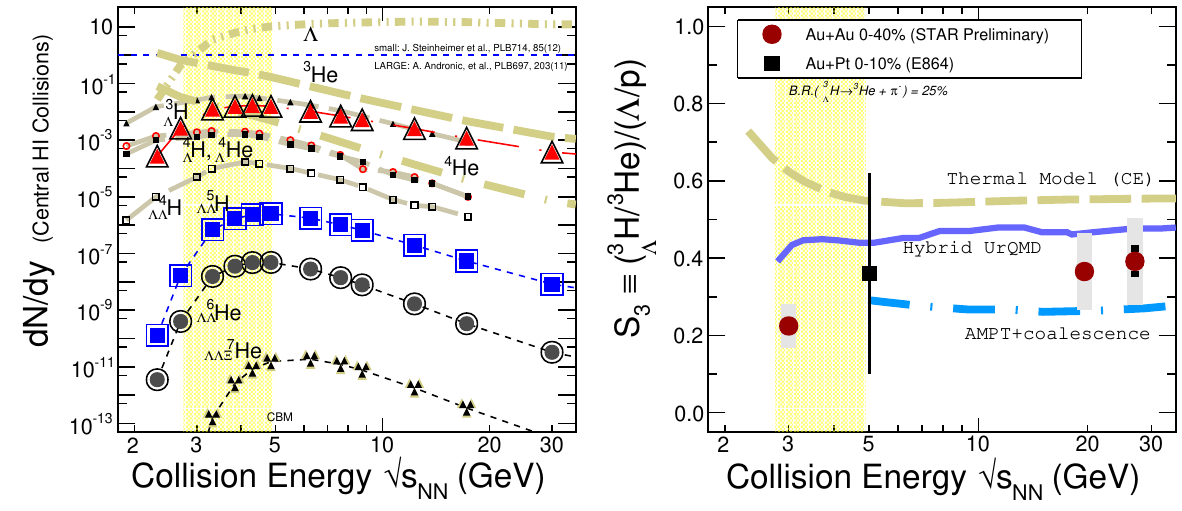} %\vspace{-1.25cm}
\linespread{1.0}\selectfont{} 
\caption{Left panel: Yields of light nuclei and hypernuclei at mid-rapidity as a function of collision energy as obtained from a thermal model~\cite{Andronic:2010qu} (large symbols) and the hadronic transport model UrQMD~\cite{Steinheimer:2012tb} (small symbols).  Right panel: Strangeness population ratio $S_3\equiv ({}^3_{\Lambda}{\rm H}/{}^3{\rm He})/(\Lambda/p)$ as a function of collision energy. Model calculations from a thermal model (dashed gold line), the hybrid UrQMD model (solid purple line), and the AMPT model with coalescence (dot-dashed blue line) are also presented. The hatched region indicates the energy range covered by the CBM experiment~\cite{Friman:2011zz}. }
\label{fig31:Yields_and_S3}
\end{figure}

The STAR and ALICE experiments have used high-energy $AA$, $pA$, and $pp$ collisions to measure hyperon-hyperon ($Y$--$Y$) and hyperon-nucleon ($Y$--$N$) correlations~\cite{STAR:2014dcy,STAR:2018uho,ALICE:2019eol,ALICE:2020mfd}. By comparing to theoretical calculations of hyperon interaction models, such as those motivated by lattice QCD calculations~\cite{HALQCD:2019wsz,HALQCD:2018qyu}, we have obtained a first glimpse of the strength and the nature (attractive or repulsive) of interactions of several hyperon species. For nucleus-nucleus collisions at fixed-target energies, the high net-baryon density and, consequently, the large number of hyperons produced in the collisions provides a unique opportunity to measure these interactions with unprecedented precision. Indeed, thermal model calculations~\cite{Andronic:2010qu,Andronic:2005yp} and UrQMD simulations~\cite{Steinheimer:2012tb} predict the production of light nuclei and hypernuclei to peak around $3 \le \snn \le 10\ \txt{GeV}$ in nuclear collisions, driven by the combination of the high value of the net-baryon density reached and the strangeness production threshold; see the left panel in Fig.~\ref{fig31:Yields_and_S3}. Moreover, unlike at the top RHIC energy and at the LHC, the hadronic phase plays a more important role in the collision dynamics at the relatively low center-of-mass energies% leads to a longer hadronic phase
, allowing hyperons and nucleons to interact over a larger fraction of the collision evolution time, thus yielding stronger correlations. 

%-- Fig.4 --===========================================================

In the right panel of Fig.~\ref{fig31:Yields_and_S3}, the experimentally measured ratio $S_3\equiv\frac{{\rm_{\Lambda}^3H/^3He}}{\Lambda/p}$, in which trivial factors (such as the chemical potential and canonical effects)
cancel to reveal the strength of the hyperon-nucleon interaction, is shown along with model calculations~\cite{Steinheimer:2012tb,Zhang:2009ba,Guo:2021udq,Ivanov:2020udj}. New precision data from STAR show a gradual increase of $S_3$ as a function of $\snn$, with the value approaching the equilibrium limit at the LHC~\cite{Borissov:2022eaj}. %Similar canonical suppression with diminishing energy is also seen in simulations of small systems at LHC energies~\cite{Sun:2018mqq} \as{please double check this sentence}. 
It is interesting to note that the limiting value in the canonical calculations of $S_3$ is about 2/3, which is commonly used in modeling of $\Lambda$--$N$ interactions~\cite{Glassel:2021rod}. Conversely, in collisions at  $\snn\le 10\ \txt{GeV}$, $S_3$ is further away from the thermal limit, suggesting an effect due to probing a high net-baryon density region. In addition, yields of light nuclei and hypernuclei are potentially sensitive to multiple-baryon correlations, which constitute an exquisite test of QCD. For example, the interaction $\Lambda$--$N$--$N$ may be important for the EoS relevant to physics inside neutron stars~\cite{Lonardoni:2014bwa}.

More precise measurements are needed in order to understand the $Y$--$N$ and $Y$--$Y$ interactions at high baryon density, which have substantial implications for the inner structure of compact stars. The RHIC BES data sets for collisions in the fixed-target mode contain on the order of 100M events at each collision energy, which is not enough for these analyses (there is an exception in the case of collisions at $\snn = 3.0\ \txt{GeV}$, for which a total of 2,000M events has been collected, and for which the initial analysis presented in Fig.\ \ref{fig31:Yields_and_S3} has been performed using 240M events collected in 2018). While the peak in the hyperon and hypernuclei production around $\snn = 4.0$--$4.5\ \txt{GeV}$ suggests that this type of measurements can be pioneered at RHIC for some of the energies covered by the BES FXT program, the CBM experiment at FAIR, with its unprecedented interaction rate capability, will be able to substantially improve the obtained constraints on the hyperon and hypernuclei production and, more generally, $Y$--$N$ and $Y$--$Y$ interactions. Beyond $\Lambda$--$N$ and $\Lambda$--$\Lambda$ correlations, the CBM experiment will also carry out a program to measure the correlations of $\Xi$--$N$, $\Omega$--$N$, $\Lambda$--$\Xi$, and possibly $\Xi$--$\Xi$.

%--==============================================================================
\subsection{Dilepton Spectra to Probe First-Order Phase Transition}\label{sec:dilepton}
%-- Fig.5 --===========================================================
%\begin{wrapfigure}{r}{0.5\textwidth}
\begin{figure}[htb]
%\vspace{-0.5cm}
\begin{minipage}{0.59\textwidth}
\includegraphics[width=\columnwidth]{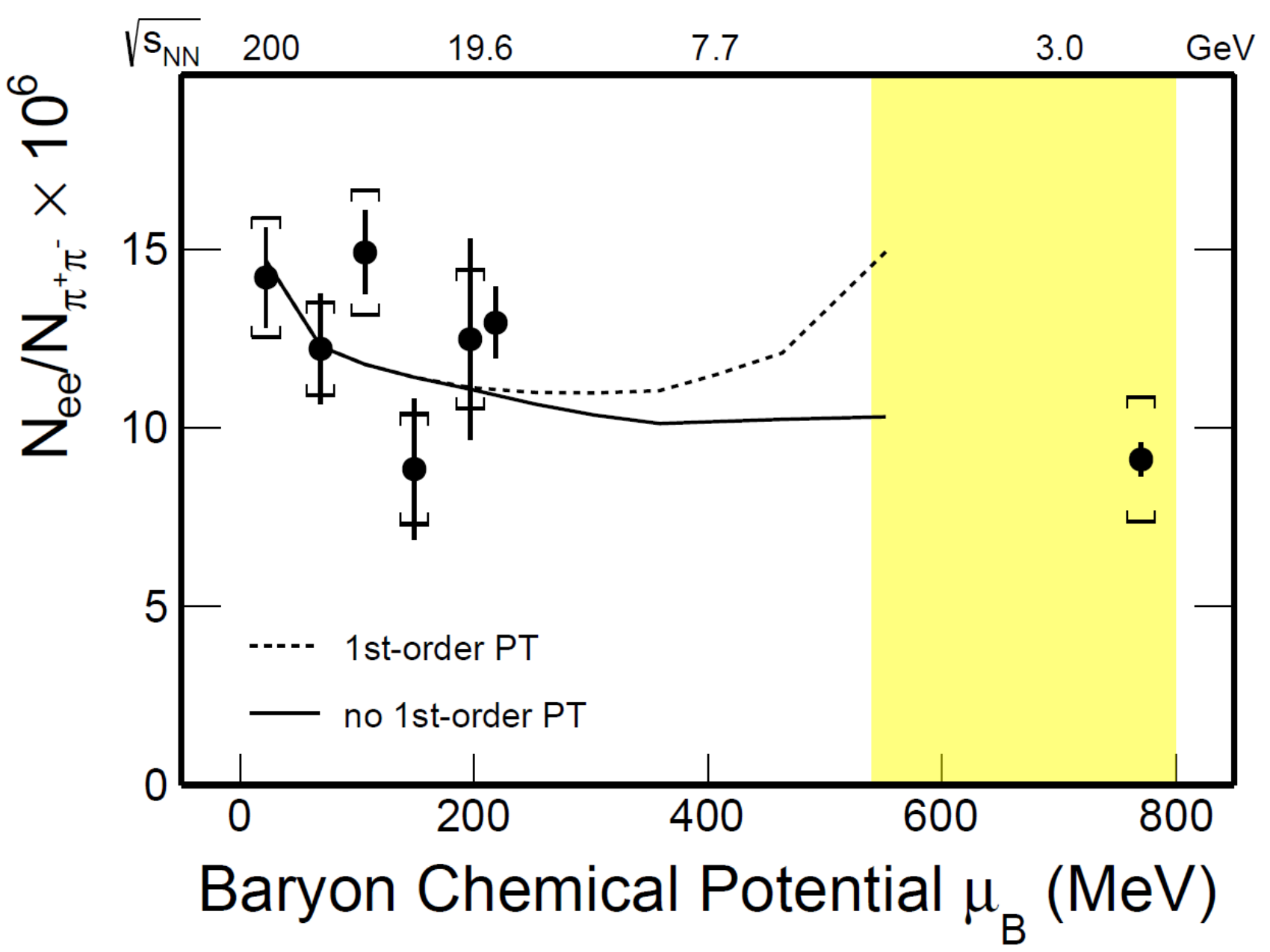}
\end{minipage}
\begin{minipage}{0.4\textwidth}
\vspace{-0.5cm}
\linespread{1.0}\selectfont{} 
\caption{The ratio of the integrated dilepton excess yield in the mass region of 0.3--0.7~GeV/$c^2$ over charged pion yield, as a function of baryon chemical potential. Data points show measurements from STAR BES-I, NA60, and HADES~\cite{STAR:2018xaj,NA60:2007lzy,HADES:2019auv}. The dashed and solid line depict model calculations with and without a first-order phase transition, respectively~\cite{Rapp:2021ee}. The yellow shaded area indicates the range of $\mu_B$ covered by the CBM experiment.}
\label{fig4:dielectron}
\end{minipage}
\end{figure}
\vspace{-0.5cm}

%\end{wrapfigure}
Real and virtual photons radiated from the hot QCD medium carry sensitive information on the  thermal and chiral properties of the medium. A combination of blue-shift--free invariant mass spectra of virtual photons (detected via dileptons) with the blue-shifted $p_T$ spectra for different masses provides particular constraints of the medium temperature as well as the collective flow throughout different phases of the evolution~\cite{NA60:2008ctj,vanHees:2007th}.
%are particularly constraining for the temperature and flow evolution
%and collective flows provide a direct (blue-shift--free) measure of the medium temperature throughout different phases of the evolution.
In consequence, they offer unique insights into the EoS of the medium in the finite temperature and density regime~\cite{Rapp:2013nxa,HADES:2019auv,Salabura:2020tou}.
Figure~\ref{fig4:dielectron} shows the integrated dilepton excess yields in the mass region of 0.3--0.7\,GeV/$c^2$ as a function of $\mu_B$ from STAR, NA60, and HADES~\cite{STAR:2018xaj,NA60:2007lzy,HADES:2019auv}. 

A recent theoretical study has shown that a softening of the EoS due to a first-order phase transition can result in an increase of the low-mass dilepton yield relative to a cross-over scenario~\cite{Seck:2020qbx}; for an example, see the dashed line in Fig.~\ref{fig4:dielectron}. At the same time, the slope parameter of the dielectron excess mass spectra, as a measure of the medium temperature, may exhibit a distinct sharp change due to a first-order phase transition~\cite{Rapp:2021ee}. Therefore, if the phase transition between hadrons and QGP occurs in the energy range $\snn=2.4$--$10\ \txt{GeV}$ (corresponding to $800 \ge \mu_B \ge 400\ \txt{MeV}$), a systematic measurement of the excitation function of dielectron production will provide an excellent opportunity to probe the first-order phase transition line,
as well as to conduct studies of the relationship between chiral symmetry
restoration and the onset of deconfinement in this regime, including the appearance of quarkyonic matter \cite{Glozman:2022lda}. 
The CBM experiment at FAIR has dedicated apparatus for both di-electron and di-muon measurements. Together with three order of magnitude higher interaction rates compared to current experiments, %HADES and STAR BES-II, 
the CBM experiment is expected to offer unprecedented precision on dilepton observables. Moreover, the conditions achieved in heavy-ion collisions at HADES, BES-II FXT, and the future CBM experiment at FAIR are
comparable to those encountered in the simulations of neutron star mergers at early times~\cite{HADES:2019auv,Most:2022wgo}, and thus of fundamental significance for nuclear astrophysics.

%--==============================================================================
\subsection{Global Polarization and Spin Alignment to Probe Vorticity Field}

\begin{figure}[htb]
\begin{minipage}{0.59\textwidth}
\includegraphics[width=0.9\columnwidth,trim={0 0.09cm 0 2.1cm},clip]{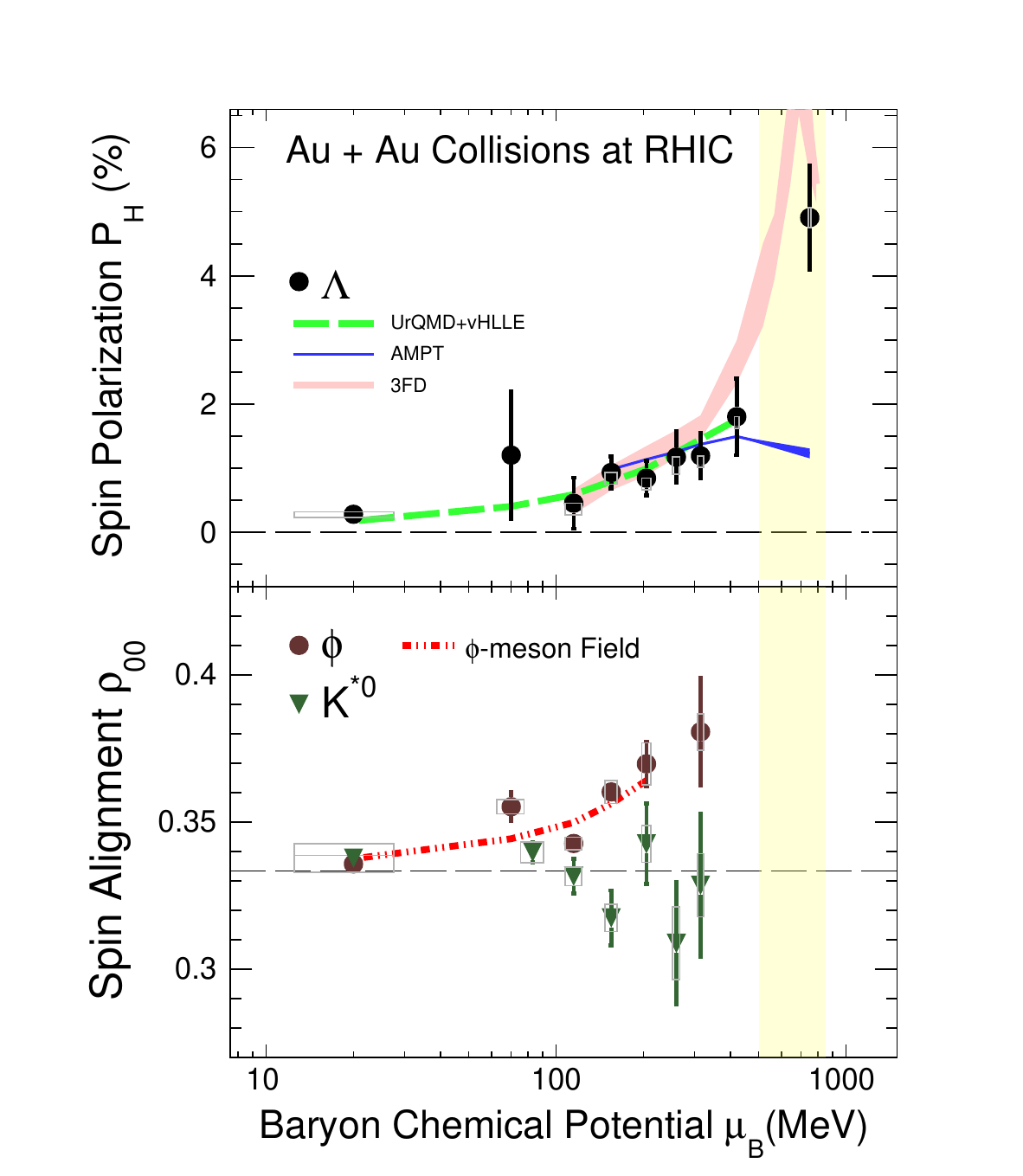} %\vspace{-.25cm}
\end{minipage}
\begin{minipage}{0.4\textwidth}
\linespread{1.0}\selectfont{} 
\caption{Top panel: $\Lambda$ spin polarization measurements~\cite{STAR:2017ckg,STAR:2021beb} (filled circles) as a function of $\mu_B$, together with model calculations by UrQMD + viscous hydrodynamics (UrQMD-vHLLE)~\cite{Karpenko:2016jyx} (dashed green line), AMPT~\cite{Guo:2021udq,Sun:2017xhx} (solid blue line), and three-fluid dynamic model 3FD (pink band)~\cite{Ivanov:2020udj}. \\
Bottom panel: vector meson $\phi$ (filled circles) and $K^{*0}$ (triangles) spin alignment measurements~\cite{STAR:2022fan} as a function of $\mu_B$, together with model calculation of the $\phi$ meson field~\cite{Sheng:2019kmk} (dot-dashed red line). Systematic uncertainties are shown by a rectangular box at each data point.}
\label{fig5:S3}
\end{minipage}
\end{figure}

The hyperon ($\Lambda$) global polarization and vector-meson ($\phi$ and $K^{*0}$) spin alignment along the direction of angular momentum in non-central heavy-ion collisions have been measured over a wide range of collision energies~\cite{STAR:2017ckg,STAR:2021beb,STAR:2022fan}; see Fig.~\ref{fig5:S3}. These results compare well with model calculations~\cite{Karpenko:2016jyx,Jiang:2016woz,Guo:2021udq,Sun:2017xhx,Ivanov:2020udj} based on rotational polarization of microscopic particle spin in a vortical fluid, suggesting the presence of a strong vorticity field. A notable trend in both the $\Lambda$ polarization and $\phi$ spin alignment measurements %(bottom panel), as seen in Fig.~\ref{fig5:S3}, 
is their rapid increase towards low beam energies where $\mu_B$ becomes large.
Is this due to a stronger baryon stopping resulting in an increased  vorticity field, as suggested by model studies~\cite{Karpenko:2016jyx,Jiang:2016woz,Guo:2021udq,Sun:2017xhx,Ivanov:2020udj}?
Is the strong $\phi$ spin alignment a consequence of the proposed $\phi$ mean field in the nuclear medium, as put forward in Ref.~\cite{Sheng:2019kmk}?
Could the vector meson mean field be connected to high baryon density and a strong vorticity field, as suggested by recent calculations~\cite{Zhang:2018ome}?

%-- Fig.5 --===========================================================
%\begin{wrapfigure}{r}{0.5\textwidth}
%\vspace{-1.0cm}
%\end{wrapfigure}

Answering these questions requires precision measurements at large $\mu_B$. The CBM experiment at FAIR, due to its unprecedented interaction rates and the capability of performing a clean $\phi$ reconstruction via both hadronic ($K^+K^-$) and dilepton channels, will provide a unique opportunity to address these questions, offering fresh insights into the structure and spin transport in dense nuclear matter in the presence of strong vorticity fields.

\subsection{Connections to Nuclear Astrophysics}

Understanding high-energy nuclear collisions and astrophysical phenomena, such as supernova explosions or neutron-star collisions, requires a robust knowledge of matter at high densities~\cite{Huth:2021bsp}. While an EoS that is soft at low densities  is needed to fit within the LIGO gravitational wave constraints, a steep rise in the speed of sound is needed above saturation density to produce recently observed $M\sim 2 M_\odot$ neutron stars. In fact, if mysterious heavy and compact objects as massive as GW190814 \cite{Abbott:2020khf} are confirmed to be neutron stars, then the speed of sound may reach nearly the causal limit \cite{Tan:2020ics}.

New methods are being developed \cite{Oliinychenko:2022uvy} to connect the neutron star EoS to that extracted from heavy-ion collisions. Future experimental data resolving open questions outlined in Sec.\ \ref{sec:collect} could provide important constraints on the neutron star EoS at $T=0$ or the $T>0$ EoS for binary neutron star mergers. 
If  a critical point and the corresponding first-order phase transition were found in  heavy-ion collisions (see Sec.\ \ref{sec:collect}, Sec.\ \ref{sec:kurtosis}, and Sec.\ \ref{sec:dilepton}), then the consequences of this first-order phase transition could be seen in neutron star mergers \cite{Most:2018eaw,Bauswein:2018bma}, in core collapse supernovae \cite{Zha:2020gjw}, or in the existence of mass twins \cite{Alford:2013aca}. 
Moreover, while microscopic models of the neutron star EoS often predict strange baryons to appear at large densities \cite{Schaffner:1995th,Dexheimer:2009hi}, they struggle to reconcile the softening hyperons cause with observational data (which is known as the hyperon puzzle).  
Measurements of strange baryon interactions as envisaged in Sec.\ \ref{sec:hyper}, could provide crucial insight into this puzzle \cite{Ghosh:2022lam}. 
%One final synergy comes from the importance of out-of-equilibrium effects within neutron star mergers \cite{Alford:2017rxf,Most:2021zvc,Most:2022yhe} that draws on knowledge from fluid dynamics developed in the context of heavy-ion collisions.  Precise measurements of collective flow at low $\snn$ will further improve that knowledge as well.

Heavy-ion collisions have given rise to an experimentally-driven field that has been crucial to the theoretical formulation of far-from-equilibrium relativistic fluid dynamics \cite{Florkowski:2010cf,Martinez:2010sc,Schenke:2010rr,Denicol:2012cn,Bazow:2013ifa,Alqahtani:2017mhy,Bemfica:2017wps,Kovtun:2019hdm} and simulations that incorporate shear and bulk viscosity \cite{Luzum:2008cw,Noronha-Hostler:2014dqa,Ryu:2015vwa,Alqahtani:2017tnq} as well as charge diffusion \cite{Denicol:2018wdp,Fotakis:2019nbq}. Furthermore, the mathematical formalism and theoretical tools developed to explore heavy-ion collisions have recently been applied to neutron star mergers, where  weak interactions lead to an effective bulk viscosity \cite{Alford:2017rxf}, opening up entirely new fields of research to explore \cite{Most:2021zvc,Most:2022yhe}. Following this parallel, one wonders whether the development of critical fluctuations in a relativistic system \cite{An:2021wof} could have consequences for neutron star mergers, especially given theoretical predictions for a low-temperature critical point induced by the axial anomaly in dense QCD matter \cite{Hatsuda:2006ps}.

%%%%%%%%%%%%%%%%%%%%%%%%%%%%%%%%%%%%%
\section{Scientific Cases Beyond RHIC BES: CBM experiment at FAIR}

%-- Fig.6 --===========================================================
\begin{figure}[htb]
\begin{minipage}{0.625\textwidth}
\includegraphics[width=\columnwidth]{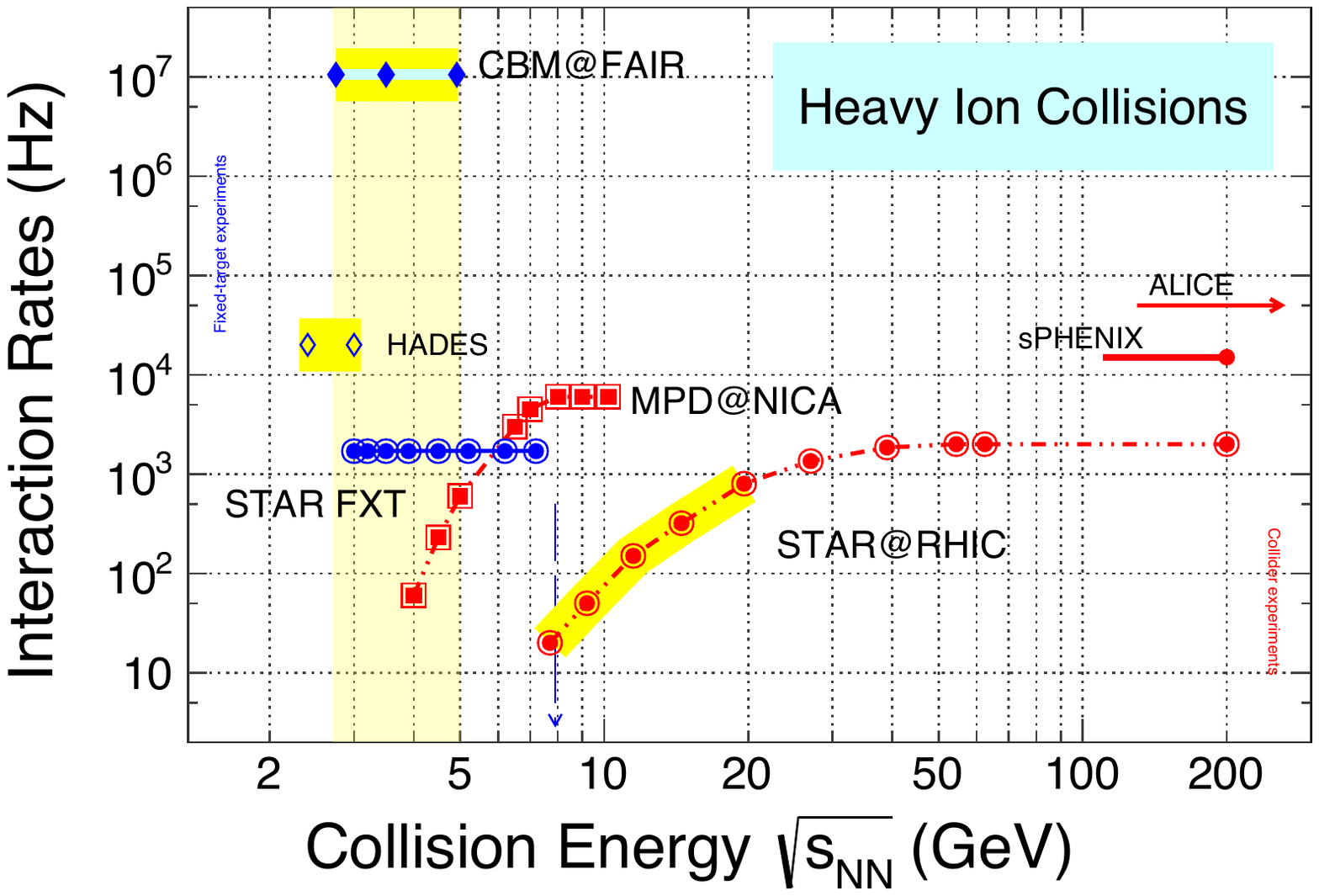} 
\end{minipage}
%\linespread{1.0}\selectfont{} 
%\vspace{-5.cm}
\begin{minipage}{0.36\textwidth}
\linespread{1.0}\selectfont{} 
\caption{Collision rates as a function of $\snn$  for collider experiments in red, and fixed-target (FXT) experiments in blue. Compared to the existing collider experiments, more than four orders of magnitude improvement in collision rates can be achieved with the future CBM experiment at FAIR. %\as{We always write that CBM starts at $\snn=2.9\ \txt{GeV}$, but here it starts around 2. Resolve the discrepancy?}
}
\label{fig6:collisionRates}
\end{minipage}\end{figure}

New accelerator facilities are being designed to have the highest possible luminosity. The heavy-ion collision rates for some of the current and future facilities and upgrades are summarized in Fig.~\ref{fig6:collisionRates}. 
The NA61/SHINE experiment is an ongoing experiment at the Super Proton Synchrotron (SPS) at CERN, investigating hadron production in collisions of beam particles (pions, protons, and beryllium, argon, and xenon ions) with a variety of fixed nuclear targets. The current program will continue until the end of 2024, and a program beyond that is being discussed. The NA60$^+$ experiment is planned as an upgrade to the NA60 experiment at the CERN SPS, and it will study dilepton and heavy-quark production in nucleus-nucleus and proton-nucleus collisions with collision energies of $\snn = 6$--$17.3\ \txt{GeV}$, with data taking aimed to start around 2029. The MPD experiment at NICA will offer a smooth connection between fixed target experiments at lower energies and collider experiments at higher energies, while the ongoing HADES experiment at GSI, operating at relatively low beam energies $\snn \approx 2.0$--$2.4\ \txt{GeV}$, allows one to study subthreshold strangeness production, virtual photon emission, and particle flow and its anisotropies for a variety of colliding ion species. At even lower collision energies, experiments at RIKEN ($\snn \approx 2.04\ \txt{GeV}$ for stable heavy ions and $\snn \approx 2.01\ \txt{GeV}$ for radioactive beams) and the Facility for Rare Ion Beams (FRIB) ($\snn \approx 1.97\ \txt{GeV}$ for radioactive beams) probe nuclear matter at and around saturation density, focusing in particular on the properties of exotic nuclei and the density-dependence of the symmetry energy.

The CBM experiment at FAIR will provide an unprecedented interaction rate at high net-baryon density, with the capability of taking data at event rates up to 10~MHz~\cite{CBM:2016kpk,Friman:2011zz}. This is more than four orders of magnitude higher than what was previously possible over a similar energy range, and it will allow unique precision measurements of crucial observables. FAIR also has the ability to collide a variety of nuclei, including isobaric species with varying isospin content, allowing a precise determination of various properties of dense QCD matter. The first phase of the CBM experiment will use the heavy-ion synchrotron SIS100 to cover the center-of-mass energy range $2.9 \leq \snn\leq4.9\ \txt{GeV}$, corresponding to baryon chemical potentials $800\ge \mu_B \ge 500$~MeV (see Table~\ref{tab1} for a comparison with the BES energies). 
The STAR BES-II data analyses are ongoing and expected to yield further insights on the nuclear phase diagram, particularly on the critical point search. However, even if the critical point is discovered in BES-II, the lower energy range covered by FAIR will be essential for the understanding of the critical point (or rather a critical region because of inevitable smearing effects) and the associated first-order phase boundary. If, on the other hand, the critical point is not evident from BES-II, the FAIR energy range will be critical to complete the search.
With respect to both the rapidity coverage and event statistics, the CBM experiment will allow to build on and improve a number of measurements in the baryon-rich region, pioneered by the RHIC BES program.
% should add HADES, FRIB,

Cost-wise, the US-CBM team would require a modest support, which would include support for student participation and for an upgrade of the Si-pixel-based vertex detector in CBM. %To enable the new initiative over the next 5-10 years.%, a fraction of the current research effort %of the author groups (to be defined later) will be redirected \as{there's something wrong with this sentence}. 
Because the culmination of the RHIC-BES physics program, especially in the fixed-target mode, will likely take place around the start of the CBM program, members of the RHIC community currently involved in the BES program will be able to seamlessly transition to the CBM program. The continued US participation in experiments probing high-density nuclear matter will not only make US-led measurements of key properties of dense nuclear matter possible, but will also support and preserve the US expertise in the field. The latter, in particular, may be of significant strategic value given the proposed $400\ \txt{MeV}/u$ upgrade of the FRIB linear accelerator (FRIB400, corresponding to $\snn \approx 2.06\ \txt{GeV}$).

The discovery potential of the CBM experiment includes, but is not limited to, probing the first-order phase transition boundary, locating the QCD critical point, measuring multi-strange hypernuclei, and constraining the nuclear matter EoS at high baryon density. Among other  observables, the following key measurements will be carried out by CBM:
\begin{enumerate}
    \item differential collective flow of protons, pions, and deuterons to study the symmetric and asymmetric dense nuclear matter equation of state;
    \item higher-order proton cumulants and cumulant ratios to search for the QCD critical point and study medium properties at high baryon densities;
    \item hyperon correlations and (multi)hypernuclei to study hyperon-nucleon and hyperon-hyperon interactions, with impacts for the QCD phase diagram and our understanding of the inner structure of compact stars;
    \item dilepton spectra and collective flow to characterize the medium temperature, study the chiral symmetry restoration and search for the first-order phase boundary in the high baryon density region; and
    \item global polarization and spin alignment at extreme baryon densities to study vorticity and spin transport.
\end{enumerate}
These measurements reflect some of the most fundamental questions, and are key to our understanding of the phases of nuclear matter, including the inner structure of astronomical objects. There is no doubt that the proposed CBM physics program, building on the success of the RHIC BES program, will be essential to the understanding of QCD matter at the highest baryon densities achievable in laboratory.

FAIR is a top-priority facility for nuclear physics in Europe. It is documented in the most recent NuPECC~\cite{NuPECC} Long Range Plan 2017 for Nuclear Physics: {\em ``FAIR is a European  flagship facility for the coming decades. This worldwide unique accelerator and experimental facility will allow for a large variety of unprecedented fore-front research in physics and applied sciences on both a microscopic and a
cosmic scale. Its multi-faceted research will deepen our knowledge of how matter and complexity emerges from the fundamental building blocks of matter and the forces among them and will open a new era in the understanding of the evolution of
our Universe and the origin of the elements."} Referring to the CBM experiment, the document continues: {\em ``The ultrarelativistic heavy-ion collision experiment CBM with its high rate capabilities permits the measurement of extremely rare probes that are essential for the understanding of strongly interacting matter at high densities~\cite{NuPECCLRP2017}."}

%%%%%%%%%%%%%%%%%%%%%%%%%%%%%%%%%%%%%
%\section*{Acknowledgments}
{\bf Acknowledgments:} We thank Drs. T. Galatyuk, N. Herrmann, K. Rajagopal, P. Senger, J. Stroth,  and colleagues from the CBM experiment for insightful discussions.

\bibliography{CBMscience} % Produces the bibliography via BibTeX.
\end{document}